\documentclass{webofc}
\usepackage[varg]{txfonts}   
\usepackage{hyperref}
\usepackage[export]{adjustbox}
\usepackage{natbib}

\usepackage{wrapfig}
\begin{document}
\title{HOSS!}
%
%

\author{
    \firstname{David} \lastname{Lawrence}\inst{1}
    \fnsep\thanks{\email{davidl@jlab.org}}
}

\institute{Thomas Jefferson National Accelerator Facility}

\abstract{%
  The Hall-D Online Skim System (HOSS) was developed to simultaneously solve two issues for the high intensity GlueX experiment. One was to parallelize the writing of raw data files to disk in order to improve bandwidth. The other was to distribute the raw data across multiple compute nodes in order to produce calibration \textit{skims} of the data online. The highly configurable system employs RDMA, RAM disks, and zeroMQ driven by Python to simultaneously store and process the full high intensity GlueX data stream.
}
\maketitle
%

\vspace{-6.75cm}
\hspace{7.5cm}
\includegraphics[height=2.5cm]{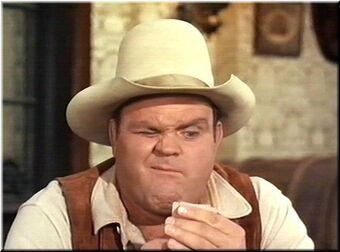}
\vspace{4.25cm}

\section{Introduction}
\label{intro}

The GlueX Experiment\cite{GlueX_NIM} is a high volume Nuclear Physics experiment installed in experimental Hall-D of the Jefferson Lab Continuous Electron Beam Accelerator Facility (aka CEBAF). The first phase of GlueX was successfully completed in 2018 where >3.5PB was acquired with DAQ system rates around 400 MB/s. For GlueX phase II, the data rate was more than doubled to approximately 1.25 GB/s. This stressed the original system developed under phase I which consisted of a single output stream written to a large capacity RAID disk server. While technically within specs for the individual components, the DAQ system exhibited instabilities when pushed to the higher rates. This motivated changes to the system to ensure stable high intensity running. Specifically, the raw data files would need to be distributed among several RAID servers in order to reduce the average rate any one server needed to support.

Another issue that came up while processing the phase I data was that it took considerable effort to extract special calibration events from the stored data files. Calibration events were typically made from special triggers for things like LED flashers used by calorimeters. The calibration events were mixed into the single output stream and were rare (<1\%) compared to the dominant \emph{physics} events. The DAQ system implementation for GlueX could not be easily changed to write separate output streams for these events directly. Thus, they needed to be extracted from the full raw data set. The process of reading the raw data from the tape archive and writing the various flavors of calibration events to files is called \emph{skimming}. The files containing only a certain type of calibration event are called \emph{skims}. An ability to generate these skim files in the counting house before the raw data ever made it to tape would save considerable time and effort in the offline. Implementing this ability in the existing DAQ system would have been difficult so it was done using a separate, new system called \emph{HOSS} for the \textbf{H}all-D \textbf{O}nline \textbf{S}kim \textbf{S}ystem. Because HOSS needed to transfer a copy of the entire 1.25 GB/s data stream to a small compute farm in the counting house, it also became a natural way to distribute the raw data files among several RAID server partitions, reducing the I/O requirements for writing to each.


\begin{wrapfigure}[8]{r}{0.2\textwidth}
    \begin{center}
        \includegraphics[width=0.18\textwidth]{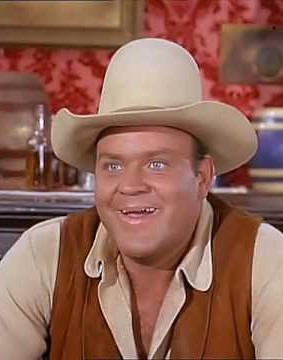}
    \end{center}
\end{wrapfigure}

The key orchestrator of HOSS is written in Python, but it relies on some key pieces of software to do the high speed network transfers and CPU-intensive computations. These components as well as details of HOSS itself are given in the following sections.

\section{RDMA Servers}
\label{sec:RDMA}

\begin{figure}[htb]
\centering
\includegraphics[width=1.0\textwidth]{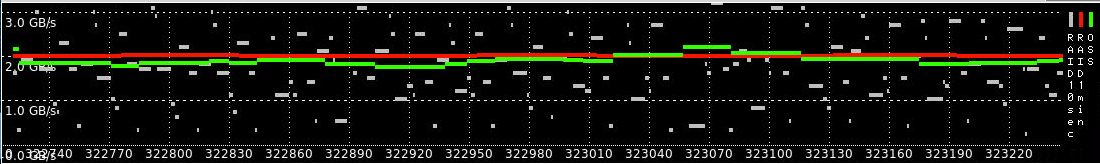}
\caption{Graph from HOSS GUI showing the integrated data rates for data received by the RDMA servers running on the RAID servers (grey and red) and the Online Skim System compute farm (green). The grey and red points are the 10 sec. and 1 minute averages. These indicate the burstiness of the system with the instantaneous rates being over 3GB/s while the longer time window indicates a 2GB/s average data transfer rate.}
\label{fig:HOSS_data_rate}       
\end{figure}

The GlueX online system includes a small farm of several($\sim16$) compute notes used for monitoring. These nodes share an infiniband network fabric with the DAQ system computers using 40Gbps(QDR) and 56Gbps(FDR) links. The hardware is all capable of using Remote Direct Memory Access (RDMA) protocols to perform high speed transfers with minimal CPU usage. Software was developed to implement reliable file transfer over RDMA with error detection. The software developed consisted of a server and client where the server was implemented as a system D service on all of the CentOS 7 computers in the counting house. In addition to RDMA, the servers also implemented a zeroMQ\cite{zeroMQ_vCHEP_format} pub-sub server over TCP/IP. This is used to check the existence of files/directories on the host as well as file sizes and checksums. This is useful for the HOSS system to check if a file has already been transferred or will overwrite an existing file before initiating a transfer. The zeroMQ interface is also used for statistics reporting to allow data rate and error rate monitoring by the HOSS system. A graphical interface is included with HOSS that displays accumulated statistics from the RDMA servers for various groups of compute nodes. In the GlueX deployment, there were two groups: The RAID servers (3 nodes) and the Online Skim System farm (9 nodes). Figure \ref{fig:HOSS_data_rate} shows a snapshot of the graph of rate vs. time from the HOSS GUI. The RDMA servers were very reliable with up times of months and each servicing >500k files (>1PB of data) during the 2020 run.

The RDMA server and client programs were compiled into a single binary. Source code for the tool can be found on GitHub here:\par
\vspace{0.5cm}
\hyperlink{https://github.com/JeffersonLab/hdrdmacp}{https://github.com/JeffersonLab/hdrdmacp}

\section{System Configuration}
\label{sec:system_configuration}

A core function of HOSS is moving files between directories on a local file system and copying the files between nodes as described in section \ref{sec:RDMA}. This function is driven by a Python script that is run on each node in the system. The script reads a common configuration file that is accessible from a network mounted directory. The configuration file format specifies rules and the list of hosts each rule should apply to. Thus, each script identifies its particular configuration by knowing its hostname and selecting the rules it should implement based on that. The two most important rules are ``stage'' and ``distribute'' and are described in the following sections.

\subsection{stage rule}
The ``stage'' rule will create hard links in one or more destination directories pointing back to a file in a source directory. Once the destination links are all created, the source directory file is unlinked. This ensures only one copy of the file ever exists on the disk and that the file is never actually physically moved once it is written. The multiple destinations allow subsequent rules applied to those directories to work independently from one another. As each rule completes, in unlinks its ``copy'' to the file. Thus, the file is removed and disk space recovered only when the final hard link is removed.

Files appearing in the source directory are only linked when there are no open file descriptors for it on the system. This ensures files do not progress through the HOSS system while they are still being written to. This is done by running the \emph{sudo /usr/sbin/lsof} command as a subprocess explicitly giving it the names of any file seen in the source directory. The subprocess is run using the \emph{sudo} command so that it can see open file descriptors held by any user, including root. This is needed since some files will be written by the RDMA server program(section \ref{sec:RDMA}) which is run as a system service and therefore as root. To avoid general admin-level access for the operations account running HOSS, a special configuration file was added to the \emph{/etc/sudoers.d} configuration to allow running that one command by that one specific account. 

\begin{figure}[htb]
\centering
\includegraphics[width=0.5\textwidth,angle=90,trim={4.5cm 0 2.5cm 0},clip]{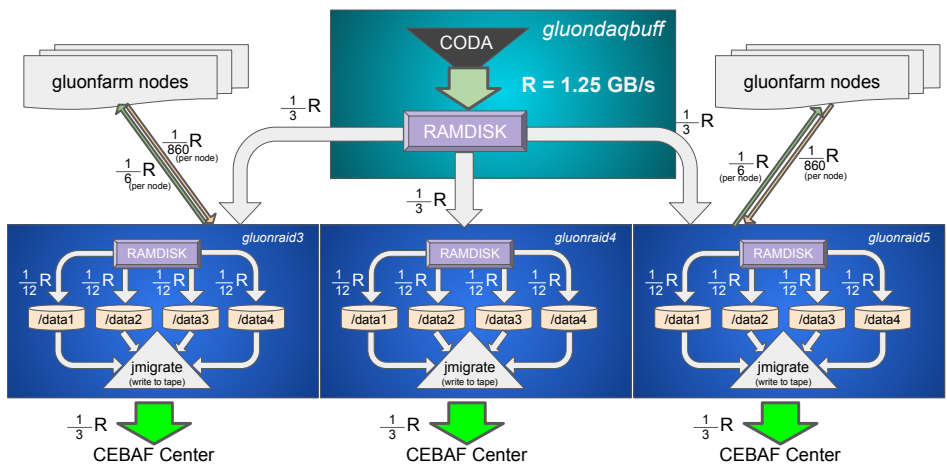}
\caption{Illustration of how HOSS is configured for GlueX phase II high intensity running. ``CODA'' is the DAQ system that is configured to write data to a RAM disk. HOSS watches specific directories for files with no open file descriptors and then moves them through the system.}
\label{fig:HOSS_diagram}       
\end{figure}

\subsection{distribute rule}
The ``distribute'' rule is used to either distribute a file from a local source directory to one of several destinations, or run a process on a file. If moving a file, the destination may be local or remote. The same mechanism used to ensure only complete files are moved in the ``stage'' rule is used here. The HOSS system recognizes if the source and destination are on the same node and will use a simple ``cp'' command in that case. Otherwise it will run the external RDMA copy tool as a subprocess.

Figure \ref{fig:HOSS_diagram} shows a diagram of how the HOSS system was configured. As illustrated in the figure, HOSS implemented heavy use of RAM disks to move the raw data around, avoiding writing to spinning disks until finally landing on the RAID servers. This results in only 1/12 of the average total data rate being written to any particular RAID partition in this particular configuration.

\section{Online Calibration skims}

\begin{wrapfigure}{l}{0.2\textwidth}
    \begin{center}
        \includegraphics[width=0.18\textwidth]{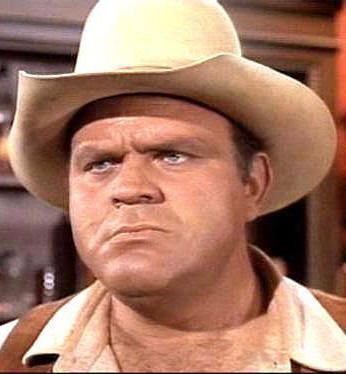}
    \end{center}
\end{wrapfigure}

A software pipeline existed for creating skim files prior to HOSS development. This included the standard GlueX reconstruction software based on JANA\cite{Lawrence_2008}\cite{Lawrence_2010}. The skim software itself is non-trivial due to the data being in an \emph{entangled} format. Entangled meaning that the data format as written by the DAQ system has pieces of 40 different events interleaved inside of a single block. Since the skim system needs to write out single events, the data must be disentangled first to create in-memory objects representing the individual events. Individual events that are to be written out must then be serialized into the output stream as single events. The disentangling itself is where the bulk of the CPU cycles are spent in the skim jobs. So much so that the available farm would have been unable to keep up with the full high intensity data rate. The issue was addressed by inserting an additional program into the skim pipeline that would filter out all entangled event blocks where none of the 40 events it contained was a calibration event. Because the calibration events represented << 1/40 of the total events, this avoided disentangling a large portion of the data stream resulting in a x4 speedup of the overall skim pipeline.

\begin{figure}[htb]
\centering
\includegraphics[width=1.0\textwidth,clip]{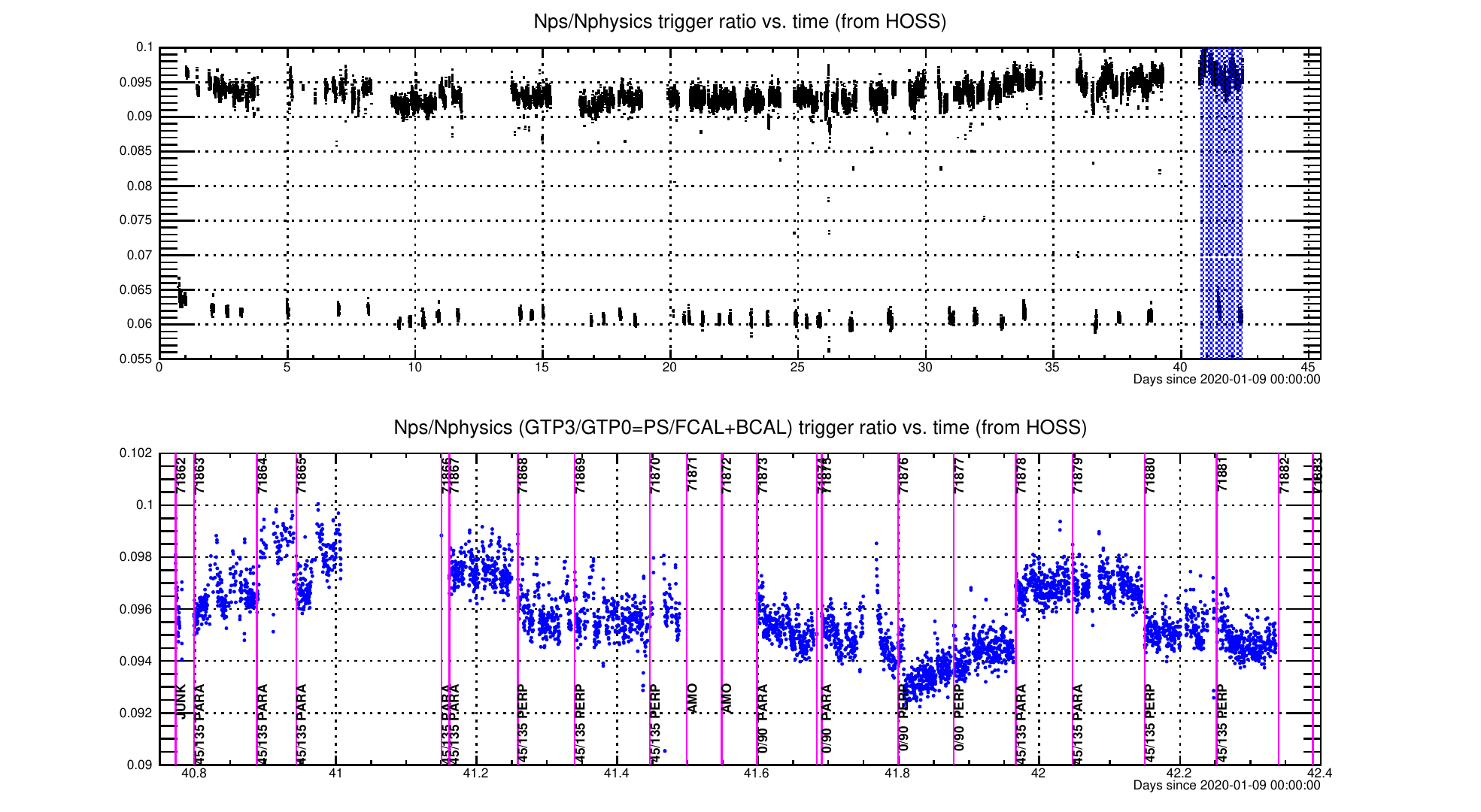}
\caption{Plots showing the ratio of pair spectrometer triggered events to the main physics trigger as a function of time using data from the HOSS DB. The top plot spans 45 days while the bottom plot shows a zoomed in view of the shaded region of the top plot. A clear dependency can be seen on the orientation of the diamond radiator (separated by the magenta lines) as well as drifts over time for a fixed radiator. This was ultimately determined to be due drifts in the beam position.}
\label{fig:trig_ratio_vs_time} 
\end{figure}

An additional feature built into the new intermediate skim process is that it was able to gather statistics on the various event/trigger types in each raw data file. Even for blocks of events that were not marked for disentangling by the next stage of the pipeline. These statistics were entered into the HOSS DB along with other metadata from the file such as the first and last event numbers. This database has proven a useful byproduct for diagnostic purposes. Figure \ref{fig:trig_ratio_vs_time} shows a plot of the ratio of the number of pair spectrometer triggered events to the number of events from the main physics trigger. The pair spectrometer is used to monitor the flux of the linearly polarized photon beam. In principle, the ratio of these two triggers should be constant since they are both proportional to the beam flux. However, a clear correlation can be seen with the orientation of the diamond radiator that is used to generate the linearly polarized photons. Moreover, the ratio can be seen to drift over time even for a fixed orientation. This type of plot with this granularity was not available in data runs prior to the introduction of HOSS.

\section{Web-accessible Database}

\begin{figure}[htb]
\centering
\includegraphics[width=1.0\textwidth]{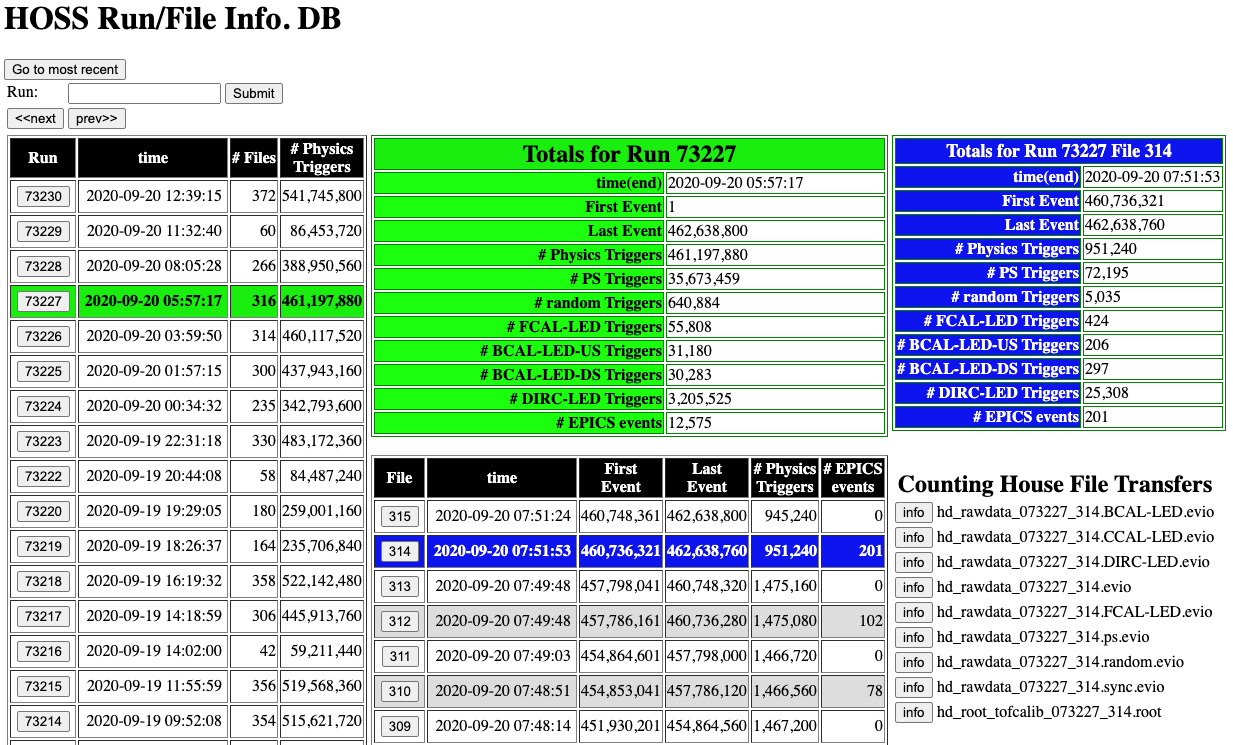}
\caption{HOSS DB Website. The HOSS system records statistics from each of the raw data files processed and each of the skim files it produces using a MySQL DB. The DB can be accessed via a dedicated webpage.}
\label{fig:HOSS_Web_DB}       
\end{figure}

Meta-data accumulated in the HOSS DB by the skim processes can be accessed by collaborators via a website. Figure \ref{fig:HOSS_Web_DB} shows a snapshot of a page indicating statistics for a single file from a 3.3hr run. Each 20GB raw data file represents about 20 seconds of beam time. Details of each stage a file passes through while in HOSS are recorded in the DB and can be accessed as well for debugging purposes (see ``info'' buttons in bottom right of figure). The website is implemented via PHP scripts accessing a replica of the MySQL DB. Replication of the DB is done in order to make the website accessible from outside the firewall protecting the internal networks in the counting house. It has the added benefit of segregating website accesses from counting house operations.

\section{Summary}

The Hall-D Online Skim System (HOSS) was built to simultaneously address a need to produce skim files online while reducing I/O bandwidth to individual RAID servers as needed by high intensity running of the GlueX experiment. The system uses Python programs applying rules defined in a configuration file making it a portable system that could be applied to other experiments. An RDMA server/client program was developed for use by the system to take care of high speed network transfers within the counting house. This can also be used as an independent tool outside of HOSS. Finally, the HOSS system records meta-data about the raw data files into a database. This has proven useful for beam diagnostics as well as offline tasks including a web-accessible interface for collaborators.
Finally, if you are curious about the pictures of the cowboy, you can lookup \emph{Hoss Cartwright} from the television show \emph{Bonanza}. A quick internet search for the word ``Hoss'' will tell you all you need to know.

\section{Acknowledgements}
This work would not have been possible without the support of the GlueX Collaboration and the JLab technical staff.

This material is based upon work supported by the U.S. Department of Energy, Office of Science, Office of Nuclear Physics under contract DE-AC05-06OR23177.

\bibliography{references}

\end{document}